\documentclass[a4paper,11pt]{article}
\usepackage{pos}
\usepackage{float}
\usepackage[caption = false]{subfig}
\title{Study of Mass Composition of Cosmic Rays with IceTop and IceCube}

\author{The IceCube Collaboration \\{\normalsize \normalfont(a complete list of authors can be found at the end of the proceedings)}}

\emailAdd{paras.koundal@kit.edu}
\emailAdd{matthias.plum@icecube.wisc.edu}
\emailAdd{julian.saffer@kit.edu}

\renewcommand{\printHeadAuthors}{Paras Koundal}

\abstract{The IceCube Neutrino Observatory is a multi-component detector at the South Pole which detects high-energy particles emerging from astrophysical events. These particles provide us with insights into the fundamental properties and behaviour of their sources. Besides its principal usage and merits in neutrino astronomy, using IceCube in conjunction with its surface array, IceTop, also makes it a unique three-dimensional cosmic-ray detector. This distinctive feature helps facilitate detailed cosmic-ray analysis in the transition region from galactic to extragalactic sources. We will present the progress made on multiple fronts to establish a framework for mass-estimation of primary cosmic rays. The first technique relies on a likelihood-based analysis of the surface signal distribution and improves upon the standard reconstruction technique. The second uses advanced methods in graph neural networks to use the full in-ice shower footprint, in addition to global shower-footprint features from IceTop. A comparison between the two methods for composition analysis as well as a possible extension of the analysis techniques for sub-PeV cosmic-ray air-showers will also be discussed.

\vspace{4mm}
{\bfseries Corresponding authors:}
Paras Koundal$^{1*}$, Matthias Plum$^{2}$, Julian Saffer$^{3}$\\
{$^{1}$ \itshape Institute for Astroparticle Physics, Karlsruhe Institute of Technology, 76021 Karlsruhe, Germany}\\
{$^{2}$ \itshape Department of Physics, Marquette University, Milwaukee, WI, 53201, USA}\\
{$^{3}$ \itshape Institute of Experimental Particle Physics, Karlsruhe Institute of Technology, 76021 Karlsruhe, Germany}\\[4mm]
$^*$ Presenter

\FullConference{37$^{\rm{th}}$ International Cosmic Ray Conference (ICRC 2021)\\
		July 12th -- 23rd, 2021\\
		Online -- Berlin, Germany}
}

\begin{document}
\maketitle
\section{Introduction}
IceCube is a cubic-kilometer astroparticle detector at the geographic South Pole, with over 5000 digital optical modules (DOMs) on 86 strings deployed in the glacial ice at depths between 1450\,m and 2450\,m \cite{aartsen2017icecube}. Reconstruction of the direction, energy and identity of penetrating particles at IceCube relies on the optical detection of Cherenkov radiation emitted in the surrounding ice. Additional to the in-ice (IC) array, the detector has a surface component, called IceTop (IT) \cite{icetop2013}. At the center of the array, the IC strings and IT stations are denser than the rest of the array. This infill region is optimized for the study of low-energy particles.\par

Besides its use as an active veto for neutrino detection, IT is utilized for cosmic-ray (CR) detection. The signal footprint measured at IT from secondary particles in CR-induced air showers allows for the reconstruction of arrival direction and energy. In addition to this, the information from coincident high-energetic muons penetrating through the ice and the deep IC array can be useful to determine the mass of a CR primary. Understanding the composition of CRs holds significant importance in our understanding of their sources and acceleration mechanisms since they are the most representative matter samples from the astrophysical sources they originate from.\par
A previous IceCube study has already demonstrated the feasibility of estimating CR energy and mass simultaneously with reconstructed IT and IC parameters using a neural network \cite{PhysRevD.100.082002}. Recent advances in machine learning have opened prospects for even faster and more detailed analyses of CRs with IceCube. Hence, here we will discuss the development on three different approaches for CR primary mass estimation. Tree-based techniques and a graph neural network (GNN) based technique are presented, which show an improvement in mass reconstruction and reduce systematic uncertainties, especially for CR primaries with intermediate mass. Also, while the previous analysis was constrained to an earlier detector configuration (79-string, 73-station,  configuration, IC-79/IT-73), we expand our methods to the complete IC86/IT81 configuration as well as to sub-PeV primary CR energies.

\section{Methods}
In the following text, we will discuss three different machine-learning (ML) and deep learning based implementations that are intended to assist in establishing a framework that can do cosmic-ray analysis on a per-event basis at IceCube. These will possibly provide us with new insights into the still uncomprehended problems in air-shower physics, in addition to reducing analysis time.
The first method is based on a random forest ensemble of high-level reconstructed air-shower parameters. The second method is based on GNNs and benefits by using the information encoded in the full-signal footprint measured in the IC, in addition to using global reconstructed information unique to each air shower.  
The third method is an alternative tree-based approach that extends to sub-PeV CRs. Since the already studied shower properties used for composition studies at IceCube do not give sufficient separation power with decreasing energy, a collection of new in-ice features is tested in the low-energy regime.\\

The baseline Monte-Carlo (MC) simulations used in these studies are simulated using the CORSIKA \cite{HeckKnappCapdevielle1998_270043064} air-shower simulation program, using FLUKA \cite{FLUKA:2006} as the low-energy hadronic interaction model and SIBYLL 2.1 \cite{Ahn_2009} as the high-energy interaction model. The proton, helium, oxygen and iron datasets were simulated with an $E^{-1}$ spectrum in the range $5.0 \leq \log_{10}(E/\mathrm{GeV}) \leq 8.0$.

\begin{figure}
\subfloat{\includegraphics[width=.4975\textwidth]{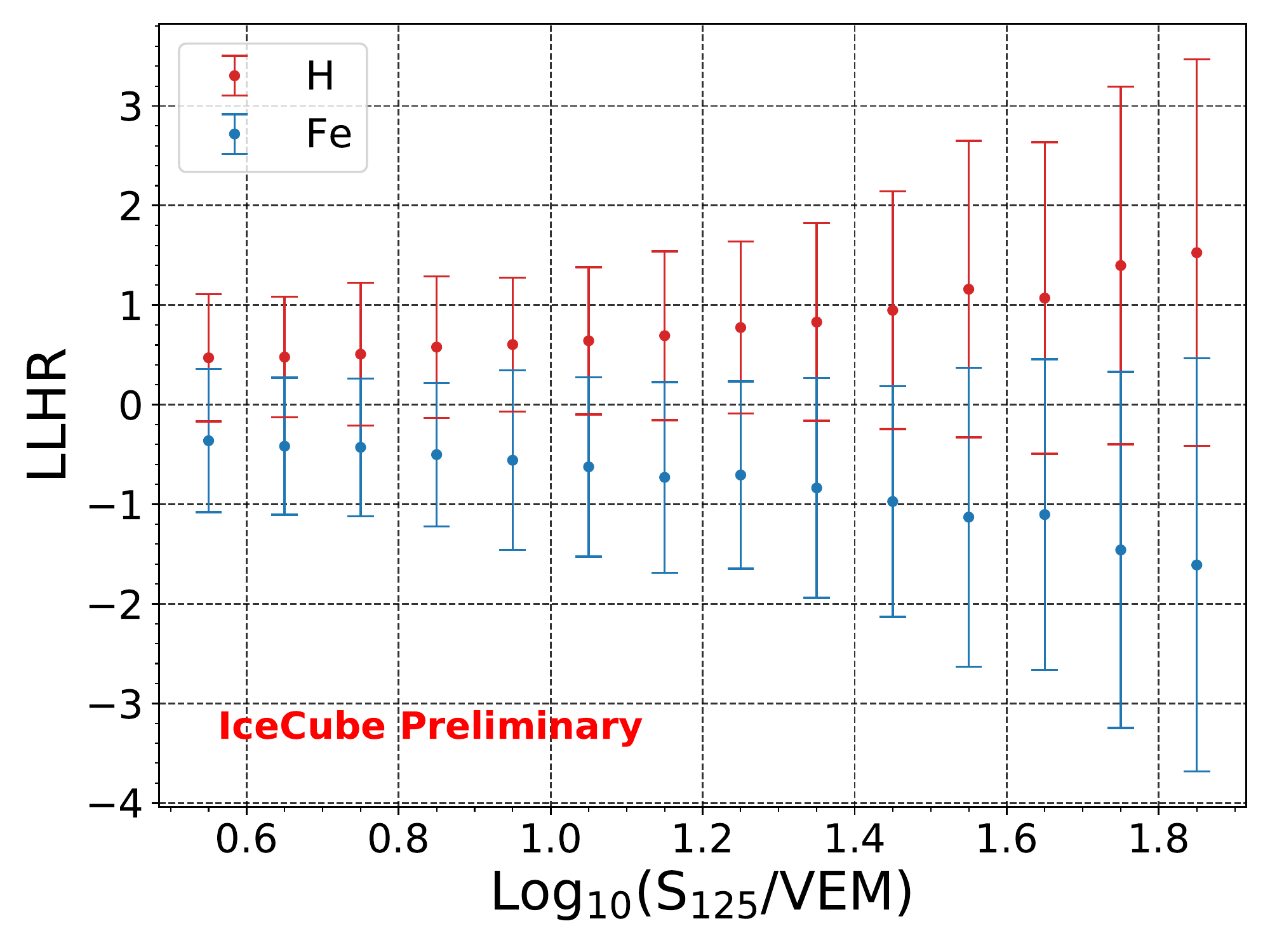}} 
\subfloat{\includegraphics[width=.5025\textwidth]{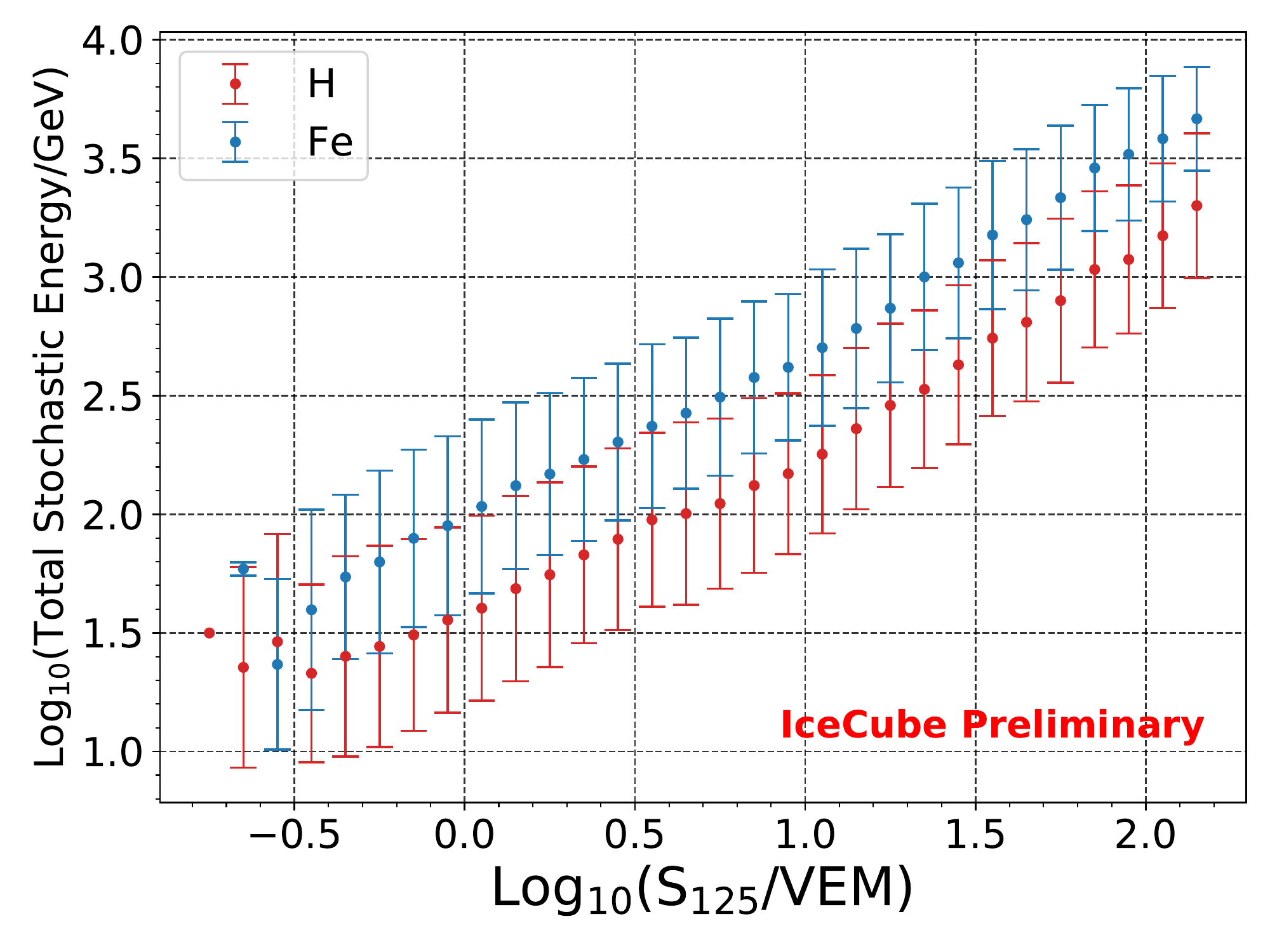}}
\caption{Composition sensitive IC86/IT81 variables (SIBYLL 2.1 \cite{Ahn_2009}, \textit{left:} log-likelihood ratio, \textit{right:} total stochastic energy) as a function of shower size, $S_{125}$, for proton and iron primaries.}
\label{mass_variables}
\end{figure}

\subsection{Based on Random Forests}
Reconstructed values are used to train a random forest (RF) ensemble to reconstruct the logarithm of the primary mass number $\ln(A)$. Two different input scenarios are compared with each other, similar to the study shown in \cite{Plum:2019lnm}, a \textit{baseline} analysis and an \textit{improved} input analysis. Scikit-learn \cite{scikit-learn} was used for pre-processing and for the computation of the RF.
Only events passing the IceTop and in-ice selection cuts mentioned in \cite{PhysRevD.100.082002} were used. 

The \textit{baseline} analysis uses the reconstructed zenith angle $(\cos(\theta))$, the logarithm of the expected shower signal $(\log_{10}(S_{125}))$ at 125\,m  distance from the shower axis in IT and the fit-value of the IC energy loss profile at a fixed slant depth of 1500\,m $(\log_{10}(\mathrm{d}E/\mathrm{d}X_{1500\,\mathrm{m}}))$. As shown in \cite{PhysRevD.100.082002}, $\log_{10}(S_{125})$ is correlated to the primary CR energy and is used as an energy estimator.

The \textit{improved} analysis uses an additional log-likelihood ratio (LLHR) value on a per-event basis, along with the remaining features mentioned. 
This LLHR value is calculated from the simulated detector response parameterization of all IceTop tanks for proton and iron MC, based on the study \cite{Pandya:201764}. The dependence of LLHR on $S_{125}$ is shown in Figure \ref{mass_variables} (\textit{left}). It shows a similar importance for mass sensitivity as is shown by IC's $\log_{10}(\mathrm{d}E/\mathrm{d}X_{1500\,\mathrm{m}})$. The additional information for the RF analysis shows an improvement on average of 10\% in the mass composition resolution in the template analysis similar to \cite{Plum:2019lnm} .
\subsubsection{RF Regressor Structure}
The same meta-parameters are used for the two RF regressors. 
In total, 500 trees are trained in this ensemble with a maximum depth of 150. The tree split criterion is the minimization of mean square error (mse). Additionally, a bootstrap method was applied and an out-of-bag sample to estimate the generalization score. All other values are kept to the default values in \cite{scikit-learn}.

\subsection{Based on Graph Neural Networks}
In the upcoming decade, IceCube Collaboration intends to enhance the current detector to the next-generation instrument, called IceCube-Gen2 \cite{Aartsen_2021}. The enhancement will add new detector-types to the current detector. This includes a high-energy radio array and a surface CR detector array, in addition to enlarging the current in-ice optical array and adding the even denser lower-energy deep-core strings around the center of the current array. These enhancements will drift the current detector geometry from the roughly hexagonal geometry to a more irregular one. A long-term analysis method implemented at IceCube needs to be flexible to these detector upgrades.\par
The current ML-based techniques implemented for cosmic-ray analysis \cite{PhysRevD.100.082002} at IceCube require a prior reconstruction of composition-relevant air-shower parameters (e.g. $\log_{10}(\mathrm{d}E/\mathrm{d}X_{1500\,\mathrm{m}})$, LLHR). A future analysis, using the entire detected IT and IC shower-footprint holds the possibility of a further improvement in mass-estimation. The utilization of the in-ice signal for neutrino analysis at IceCube Observatory is more detailed, and the implementation uses convolutional neural networks (CNNs) \cite{abbasi2021convolutional}. The successful training of a CNN requires data arranged in a fixed and uniform orthogonal grid. The current detector geometry is closer to a hexagonal geometry than orthogonal. As the strings and DOMs, which form the input pixels for the CNN in the analysis, are not uniformly distanced throughout the detector, the current CNN-based implementation has to transform the hexagonal geometry to an orthogonal one with specialized kernels (as done in \cite{abbasi2021convolutional}) and treat the geometrically distinct parts of the detector separately. With the shift to an even more irregular detector geometry with future enhancements, these transformations will possibly be inefficacious.\par
To establish a robust analysis method which uses detailed signal deposit information and also allows easier re-implementation for a future detector configuration, GNNs \cite{Wu_2021} hold a significant promise. This is primarily because GNNs allow network training using data in its most natural configuration, while also exploiting the information exchange between individual building blocks of the graph, as is true for CNNs. A previous work using GNNs has already shown improvement over CNNs for IceCube signal classification \cite{choma2018graph}. Here, we have extended the work for the composition analysis of cosmic rays at IceCube. Another ongoing work \cite{Minh:2021icrc}, is also using GNNs for the reconstruction of neutrino events at IceCube.

\subsubsection{Network Architecture}
As with the RF method, only events passing the IceTop and in-ice selection cuts mentioned in \cite{PhysRevD.100.082002} were used. 
The network is currently implemented in PyTorch \cite{paszke2019pytorch} and takes about 12 hours for a successful training run on NVIDIA Tesla V100. An advanced implementation of the current method will soon be ported to use PyTorch Geometric \cite{fey2019fast}. PyTorch Geometric is better suited for graph-based deep-learning methods and hence also provides more opportunities for experimentation.\par
The current network architecture is divided into two components. The first one is a graph-neural network based implementation. The nodes of the graph are formed by all the in-ice hit-DOMs in an event and the edges are the learned functions of spatial coordinates of DOMs. Associated to each node is a set of input features which capture the spatial coordinates of DOMs, and the charge and timing information of the measured waveform for the hit DOMs. Non-hit DOMs are padded as zeros. The second component is a fully-connected (FC) network that uses the global features associated to each event. Similarly to the previous method, it also uses $\log_{10}(S_{125}/\mathrm{VEM})$ (vertical equivalent muon = unit for quantifying charge deposit by a vertical muon in IceTop DOMs) as an energy proxy. To benefit from the in-ice information, it uses $\log_{10}(\mathrm{d}E/\mathrm{d}X_{1500\,\mathrm{m}})$. In addition to this, it also uses total energy of high-energy local-stochastic deposits in an event (standard selection - explained in \cite{PhysRevD.100.082002}). Its dependence on $\log_{10}(S_{125}/\mathrm{VEM})$ is shown in Figure \ref{mass_variables} (\textit{right}). The output from the earlier mentioned two components are then concatenated into another FC network. The full network is trained as a regression-model with the logarithmic mass of CR primary i.e. $\ln(A)$ as the expected output.

\subsection{Low-Energy Extension with Boosted Decision Trees}
Knowing the elemental composition of cosmic rays over many energy decades enables drawing better conclusions about the origin of galactic and extragalactic CRs. In order to study the entire knee region of the cosmic-ray spectrum, in particular, the overlap region between direct and indirect measurements of CRs, lowering the energy threshold of IceTop to about $10^5$\,GeV is necessary. For this purpose, a dedicated trigger has been developed that requires only two hit nearby stations in the denser infill region in contrast to five stations needed in the standard trigger. This trigger was used for a previous IceTop analysis in the energy range 250\,TeV - 10\,PeV, however the unknown composition was a significant systematic uncertainty \cite{koirala2019low}. We use a boosted decision tree (BDT) \cite{scikit-learn} with new parameters to predict the mass of CR primaries activating that low-energy trigger.\par 
Prior to this BDT, random forest regressors \cite{scikit-learn} are trained to estimate the shower core position on the surface, zenith angle and energy of a given CR event using IT information alone. The used BDT classifier model is constrained to a maximum tree depth of 2 and is trained in 4000 iteration steps with a learning rate of $0.01$. Those hyper-parameters result from 10-fold cross-validation. The input features include fit values to the in-ice energy loss profile ${\mathrm{d}E}/{\mathrm{d}X}$ at slant depths 1500\,m and 1800\,m, stochastic losses (highest, total and average in standard and strong selection) \cite{PhysRevD.100.082002} as well as the average depth of stochastic losses and the number of hit in-ice DOMs after cleaning. For most of these features $f$, $\log_{10}(f/E_\mathrm{reco})$ is used as an input, where $E_\mathrm{reco}$ is the output of the energy regressor. RFs and the BDT are each trained on proton and iron Monte-Carlo data. Only events passing the cuts mentioned in \cite{koirala2019low} have been used for training and testing. The data used with the BDT are further constrained by requiring a successful energy loss fit to the in-ice signals. In contrast to the regressive estimation of shower core position, zenith angle and energy, here the identification of CRs is performed as a binary classification task.

\section{Results}
\begin{figure}
\centering
\includegraphics[width=0.9\textwidth]{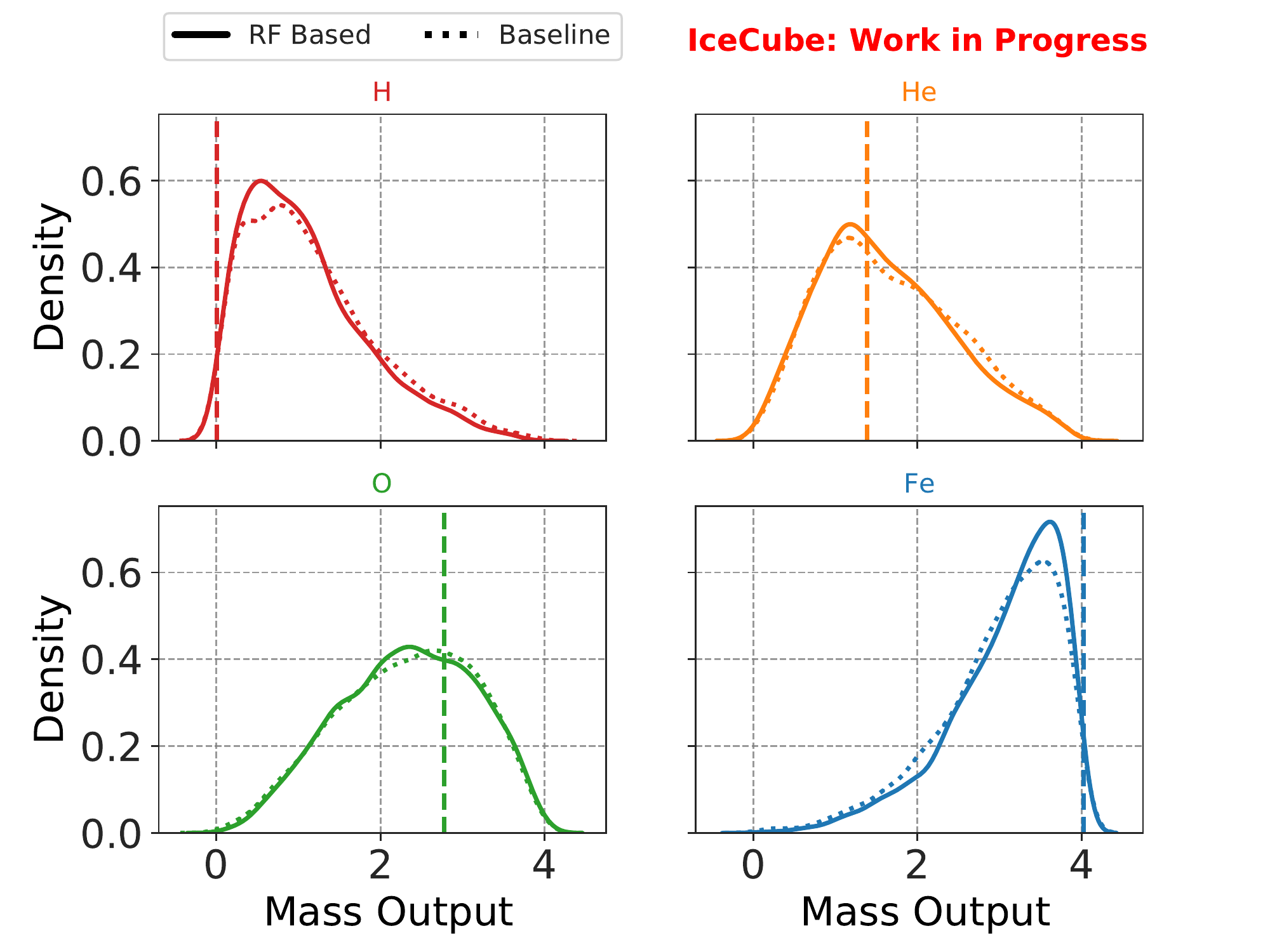}
\caption{Kernel density estimation (KDE) of the natural logarithm of
the cosmic-ray primary mass using the baseline and the improved RF-based method (vertical dashed lines: true primary mass).}
\label{fig:comparison_rft_baseline}
\end{figure}

The RF-based method benefits from the simplicity of the data-quality selection, easier model parameter selection, and the capability to determine the feature importance of the used observable parameters. As shown in Figure~\ref{fig:comparison_rft_baseline}, an improvement in the mass resolution was obtained by the RF-based method, in comparison to the baseline analysis. This improvement can primarily be attributed to the additional composition-sensitive LLHR-parameter. 

\begin{figure}
\centering
\includegraphics[width=0.8\textwidth, height = 0.6\textwidth]{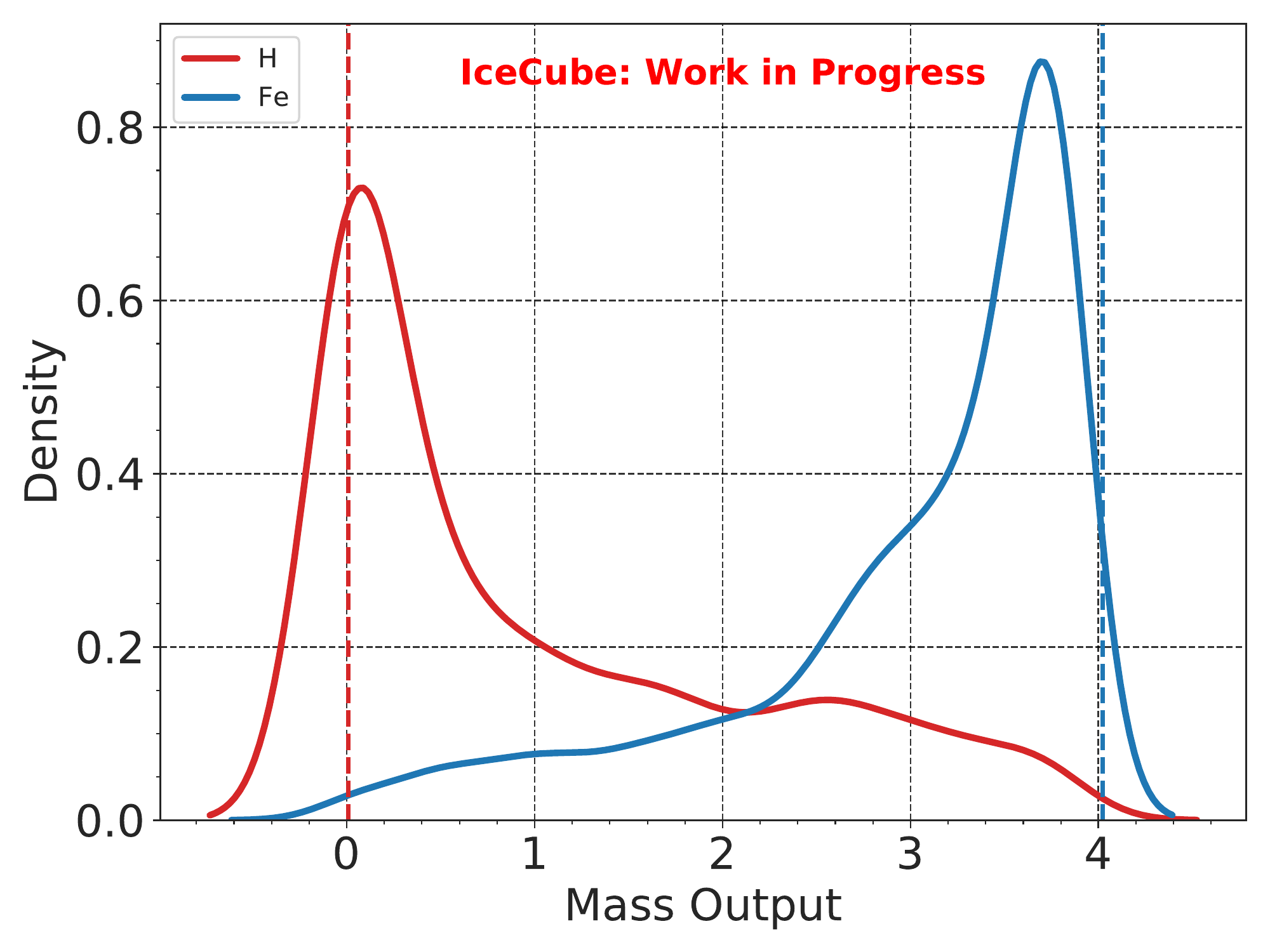}
\caption{KDE of the natural logarithm of the cosmic-ray primary mass as predicted by the GNN-based method, trained on H and Fe MC-data with a mixed composition i.e. 50\% H - 50\% Fe (vertical dashed lines: true primary mass).}
\label{fig:prelim_gnn}
\end{figure}

In the preliminary test, the GNN based technique has shown even better mass resolution and preciseness in prediction. This is shown in Figure~\ref{fig:prelim_gnn}, where the mass output by the network is centered closer to the true primary-mass for both primary types. A more detailed treatment of the observed air-shower helps in achieving better composition sensitivity than other methods. By combining high-level reconstruction information with the full detector response on a per-event basis ensures that it also benefits from already existing composition-sensitive parameters.

\begin{figure}
\subfloat{\includegraphics[clip, trim=1.5cm 0.15cm 4.6cm 0.1cm, width=.441\textwidth]{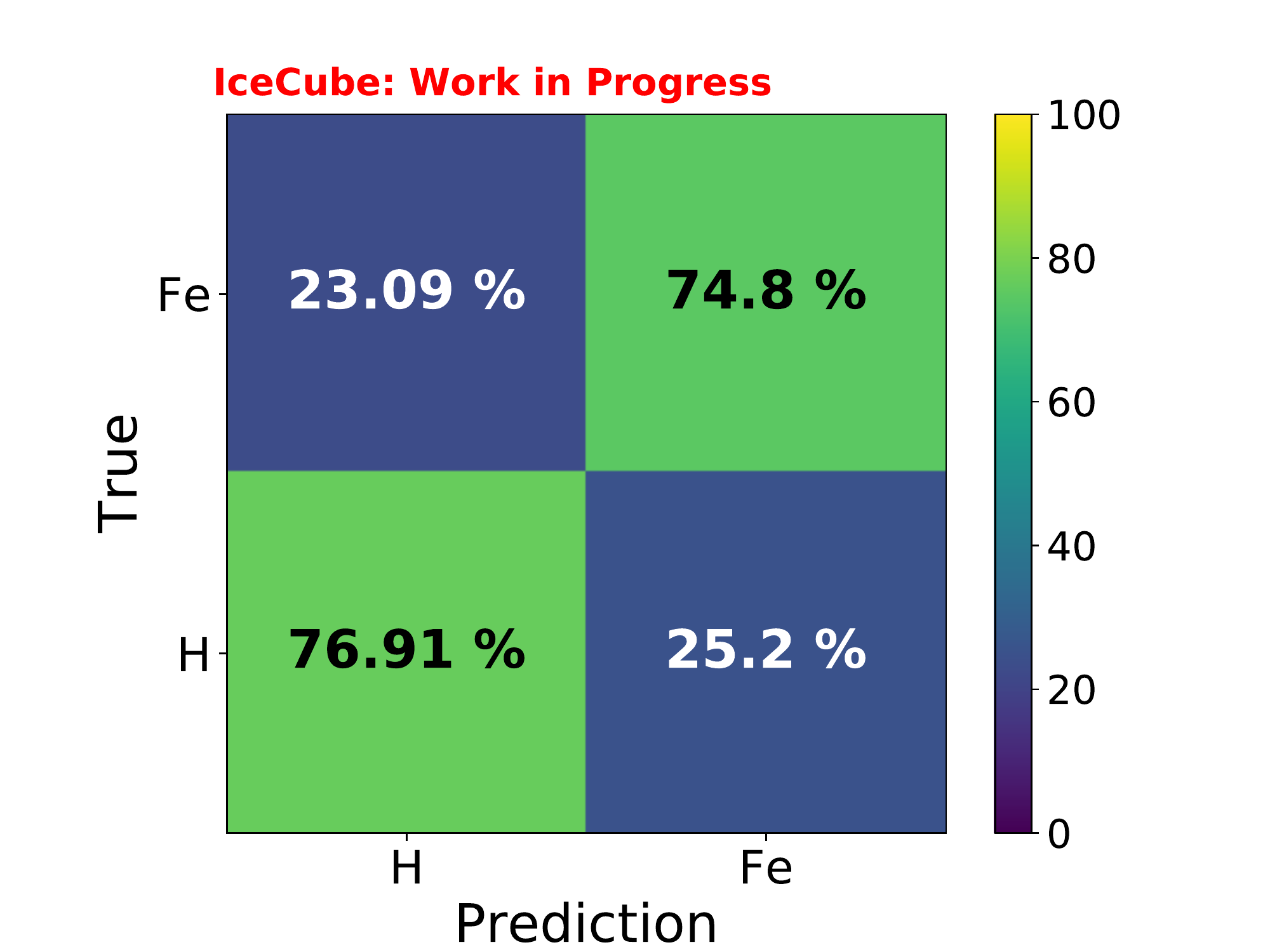}}
\subfloat{\includegraphics[clip, trim=0.4cm 0.4cm 0.4cm 0.45cm, width=.559\textwidth]{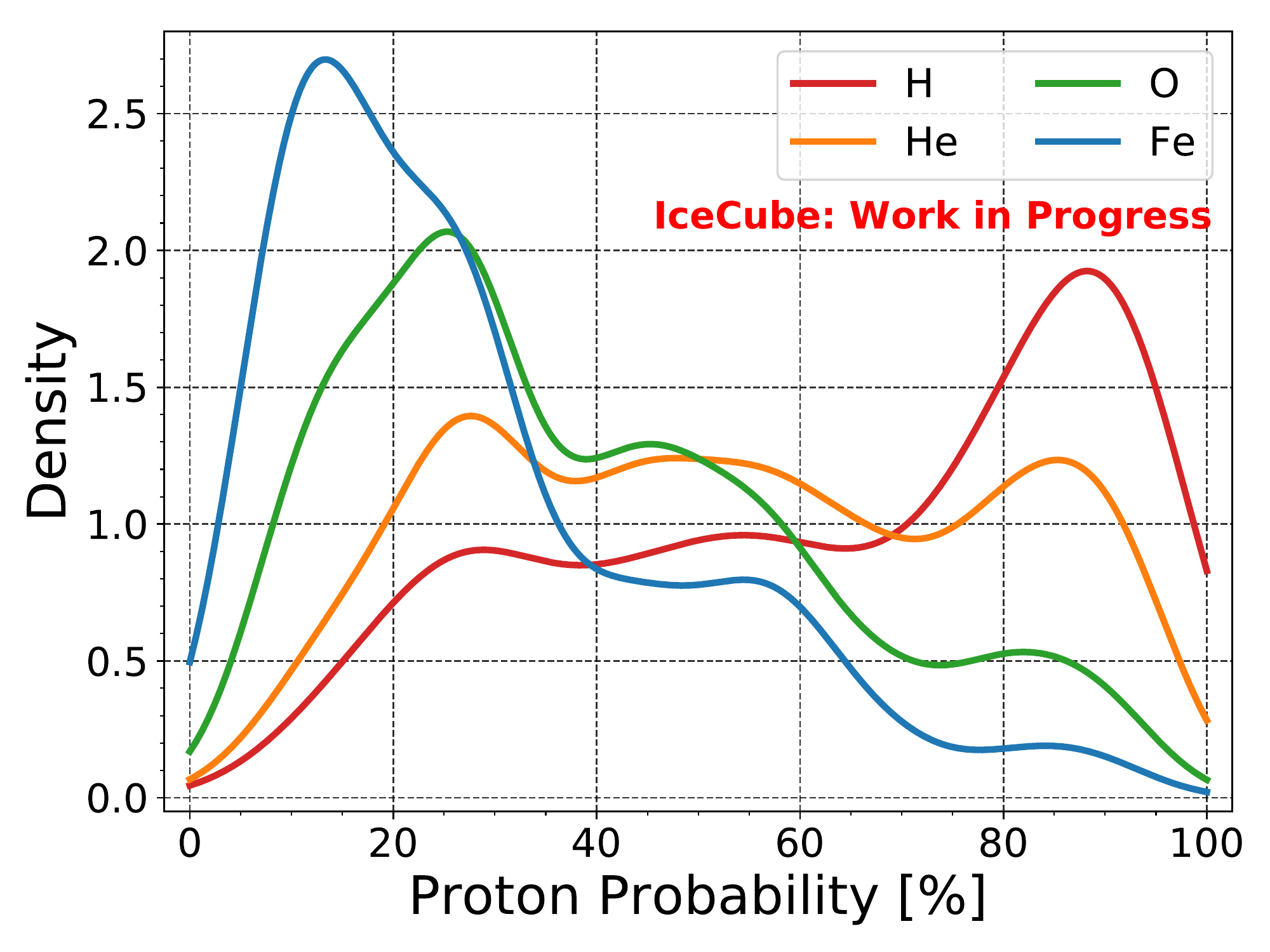}}
\caption{\textit{Left:} Confusion matrix for the prediction of proton and iron nuclei by the BDT, \textit{right:} KDE of the probabilities for proton classification for primaries proton, helium, oxygen and iron.}
\label{fig:BDT}
\end{figure}

The separation power of commonly used composition-sensitive parameters is shrinking with decreasing primary energy of CRs.
Tree-based methods, however, are able to classify cosmic-ray events with energies as low as a few hundred TeV when provided with an energy estimate from IT and combined information on in-ice energy losses. The accuracy of the assignment into the classes H and Fe is presented in a confusion matrix (Figure~\ref{fig:BDT}, \textit{left}). In order to show that the proton- and iron-trained BDT predicts the identity of a CR primary in a reasonable fashion, the model is tested on simulated data of the intermediate-mass elements, helium and oxygen. For each of the four primary types, the probabilities assigned by the BDT to a classification as proton are shown in the right panel of Figure~\ref{fig:BDT}.

\section{Conclusion and Outlook}
With the implementation of further detector types in the near future in IceCube-Gen2, the machine learning methods are crucial to derive the best physics results possible. Three different cosmic-ray mass composition studies presented here already utilize the hybrid measurements of IC and IT. All of these studies show promising results to extend and improve the mass composition sensitivity of CRs over to the full energy range, from PeV to EeV. The GNN and BDT-based approach will upgrade the analysis to all primary-types, in the immediate future. A combined future analysis will merge the best principles from all these implementations for a more detailed and precise cosmic-ray composition analysis.  

\bibliographystyle{ICRC}
\bibliography{main}

\clearpage
\section*{Full Author List: IceCube Collaboration}




\scriptsize
\noindent
R. Abbasi$^{17}$,
M. Ackermann$^{59}$,
J. Adams$^{18}$,
J. A. Aguilar$^{12}$,
M. Ahlers$^{22}$,
M. Ahrens$^{50}$,
C. Alispach$^{28}$,
A. A. Alves Jr.$^{31}$,
N. M. Amin$^{42}$,
R. An$^{14}$
% ...
\\

\noindent
% ...
$^{12}$ Universit{\'e} Libre de Bruxelles, Science Faculty CP230, B-1050 Brussels, Belgium \\
% ...
$^{14}$ Department of Physics and Laboratory for Particle Physics and Cosmology, Harvard University, Cambridge, MA 02138, USA \\
% ...
$^{17}$ Department of Physics, Loyola University Chicago, Chicago, IL 60660, USA \\
$^{18}$ Dept. of Physics and Astronomy, University of Canterbury, Private Bag 4800, Christchurch, New Zealand \\
% ...
$^{22}$ Niels Bohr Institute, University of Copenhagen, DK-2100 Copenhagen, Denmark \\
% ...
$^{28}$ D{\'e}partement de physique nucl{\'e}aire et corpusculaire, Universit{\'e} de Gen{\`e}ve, CH-1211 Gen{\`e}ve, Switzerland \\
% ...
$^{31}$ Karlsruhe Institute of Technology, Institute for Astroparticle Physics, D-76021 Karlsruhe, Germany  \\
% ...
$^{42}$ Bartol Research Institute and Dept. of Physics and Astronomy, University of Delaware, Newark, DE 19716, USA \\
% ...
$^{50}$ Oskar Klein Centre and Dept. of Physics, Stockholm University, SE-10691 Stockholm, Sweden \\
% ...
$^{59}$ DESY, D-15738 Zeuthen, Germany \\
% ...

\subsection*{Acknowledgements}

\noindent
USA {\textendash} U.S. National Science Foundation-Office of Polar Programs,
U.S. National Science Foundation-Physics Division,
U.S. National Science Foundation-EPSCoR,
Wisconsin Alumni Research Foundation,
Center for High Throughput Computing (CHTC) at the University of Wisconsin{\textendash}Madison,
Open Science Grid (OSG),
Extreme Science and Engineering Discovery Environment (XSEDE),
Frontera computing project at the Texas Advanced Computing Center,
U.S. Department of Energy-National Energy Research Scientific Computing Center,
Particle astrophysics research computing center at the University of Maryland,
Institute for Cyber-Enabled Research at Michigan State University,
and Astroparticle physics computational facility at Marquette University;
Belgium {\textendash} Funds for Scientific Research (FRS-FNRS and FWO),
FWO Odysseus and Big Science programmes,
and Belgian Federal Science Policy Office (Belspo);
Germany {\textendash} Bundesministerium f{\"u}r Bildung und Forschung (BMBF),
Deutsche Forschungsgemeinschaft (DFG),
Helmholtz Alliance for Astroparticle Physics (HAP),
Initiative and Networking Fund of the Helmholtz Association,
Deutsches Elektronen Synchrotron (DESY),
and High Performance Computing cluster of the RWTH Aachen;
Sweden {\textendash} Swedish Research Council,
Swedish Polar Research Secretariat,
Swedish National Infrastructure for Computing (SNIC),
and Knut and Alice Wallenberg Foundation;
Australia {\textendash} Australian Research Council;
Canada {\textendash} Natural Sciences and Engineering Research Council of Canada,
Calcul Qu{\'e}bec, Compute Ontario, Canada Foundation for Innovation, WestGrid, and Compute Canada;
Denmark {\textendash} Villum Fonden and Carlsberg Foundation;
New Zealand {\textendash} Marsden Fund;
Japan {\textendash} Japan Society for Promotion of Science (JSPS)
and Institute for Global Prominent Research (IGPR) of Chiba University;
Korea {\textendash} National Research Foundation of Korea (NRF);
Switzerland {\textendash} Swiss National Science Foundation (SNSF);
United Kingdom {\textendash} Department of Physics, University of Oxford.
\end{document}